\documentclass{ws-ijmpcs}

\begin{document}

\markboth{G. H. Souza \& E. Kemp \& C. Chirenti }
{Magnetized Neutron Stars}

\catchline{}{}{}{}{}
%
%%%%%%%%%%%%%%%%%%%%%%%%%%%%%%%%%%%%%%%%%%%%%%%%%%%%%%%%%%%%%%%%%%%%

\title{Magnetized Neutron Stars.}

\author{ Gibran H. de Souza}

\address{Instituto de $F\acute{\imath }sica$ \textit{Gleb Wataghin}, \textit{UNICAMP}, Cidade $Universit\acute{a}ria$ Zeferino Vaz - $Bar\tilde{a}o$ Geraldo\\
Campinas, $S\tilde{a}o$ Paulo 13083-859, 
Brazil\\
gibaifi@ifi.unicamp.br}

\author{Ernesto Kemp}

\address{Instituto de $F\acute{\imath }sica$ \textit{Gleb Wataghin}, \textit{UNICAMP}, Cidade $Universit\acute{a}ria$ Zeferino Vaz - $Bar\tilde{a}o$ Geraldo\\
Campinas, $S\tilde{a}o$ Paulo 13083-859, Brazil\\
kemp@ifi.unicamp.br}

\author{Cecilia Chirenti}

\address{Centro de $Matem\acute{a}tica$, Computa\c{c}$\tilde{a}$o e Cogni\c{c}$\tilde{a}$o, \textit{UFABC}, Av. dos Estados, 5001 - $Bang\acute{u}$,\\
Santo $Andr\acute{e}$, $S\tilde{a}o$ Paulo 09210-170, Brazil\\
cecilia.chirenti@ufabc.edu.br}

\maketitle

\begin{history}
\received{Day Month Year}
\revised{Day Month Year}
\published{Day Month Year}
\end{history}

\begin{abstract}
Here we solve numerically the relativistic Grad-Shafranov equation for a typical neutron star with 1.4 solar masses, we find the magnetic field, with both poloidal and toroidal components, inside the star and study the behavior of its field lines as a function of the ratio between the toroidal and the poloidal field.

\end{abstract}

\ccode{PACS numbers:}

\section{Introduction}	
Neutron stars are a class of compact astronomical bodies with exotic characteristics, which lie in
the range of relativistic objects together with black holes. They are remains of stellar cores that survived the  collapse of a Supernova event, and now live in a state of supra nuclear densities, with mass between 1.4 and 2.4 solar masses and radius of the order of 10km. Their average matter density easily reaches the density in the atomic nucleus \cite{bib1}. 
Recent measured periods and spin down rates of \textit{soft-gamma repeaters} \textit{(SGR)} and of \textit{anomalous X-ray pulsars }\textit{(AXP)} show that some neutron stars have a very strong and dynamical magnetic field $(∼ 10^{15} G)$ in comparison with more quiet and less active ones. These highly-magnetized neutron stars are known as \textit{magnetars}\cite{bib2}.  

Pulsars are fast rotating neutron stars with a collimated radiation flux from their magnetic poles. Due to their high rotation of few milliseconds and high magnetic field $(\lesssim 10^{10}G)$ the ions in the neutron star's surroundings are accelerated and follow the magnetic field lines until they collide with the neutron star surface at its magnetic poles, emitting in this process two radiation cones, one for each magnetic pole. When an observer is in the radiation cone path, he sees a periodic radiation pulse with the period of the neutron star's rotation\cite{bib2}.
   
Magnetars have slow periods of few seconds and a very high magnetic field $(\lesssim 10^{15}G)$ in comparison with standard neutron stars. Some believe\cite{bib2} that their slow rotation is a direct consequence of their high magnetic fields. Because of this slow rotation, \textit{magnetars} are unable to produce radiation cones like \textit{pulsars}, but a class of very energetic event is related with their bursts, the \textit{soft-gamma repeaters} (\textit{SGR}). 

According to the standard magnetar model, the reconnection of the magnetic field lines near the neutron star's crust releases a large amount of energy that excites the crust oscillation modes and this induces the production of gamma ray pulses with the same frequency \cite{bib3}. All this gamma radiation is released in an energy burst of a few milliseconds of duration.         

In this work we focus on a mathematical approach that describes the magnetic field in the neutron star's frame we model our magnetic field using \textbf{the relativistic Grad-Shafranov equation}\cite{bib4}. With this approach we are able to model the influence of the magnetic field in the crust oscillation frequency which will be our future work. In section $\textbf{2}$ we derive the relativistic Grad-Shafranov equation. In section $\textbf{3}$ we describe the methods used for solving the equation numerically. The results are shown in section $\textbf{4}$ and the final conclusions in section $\textbf{5}$.

\section{The relativistic Grad-Shafranov equation}
Working with the Maxwell equations in General Relativity we are able to describe a general static magnetic field with both poloidal and toroidal components. We start our description with a generic vector potential $A_{\mu} (r,\theta )$, which the spherical symmetry $A_{\mu }$ is a function of only $r$ and $\theta $\cite{bib4}:
\begin{equation}
 A(r,\theta )_{\mu }=(0,A_{r},A_{\theta },A_{\phi }).
 \end{equation} 
We set our background metric as a stationary and spherically symmetric metric: 
\begin{equation}
ds^{2}= -e^{\nu (r)}dt^{2} + e^{\lambda  (r)}dr^{2} + r^{2}(d\theta ^{2}+sin^{2}\theta d\phi ^{2}
), 
\end{equation}
where the functions $e^{\nu (r)}$ and $e^{\lambda (r)}$ are given by the Tolman-Oppenheimer-Volkoff equations.

We can gauge away the four-vector potential $\theta $ component by using two functions $\Lambda (r,\theta )$ and $\Sigma (r,\theta )$ such that:
\begin{equation}
\Lambda _{,\theta }=A_{\theta };
\end{equation}
and
\begin{equation}
\Sigma (r,\theta )\equiv e^{\frac{\nu -\lambda }{2}}(A_{r}-\Lambda _{,r});
\end{equation}
so that the four-vector $A_{\mu }$ becomes:
\begin{equation}
 A(r,\theta )_{\mu }=(0,e^{\frac{\lambda-\nu }{2}}\Sigma,0,A_{\phi } ).
 \end{equation} 
Imposing the zero force condition and using the Maxwell equations we find that both functions $\Sigma $ and $A_{\phi }$ can be written as function of a new function $a(r,\theta )$\cite{bib4} as:
\begin{equation}
\Sigma =\zeta a,
\end{equation}
\begin{equation}
A_{\phi }=\sin\theta a_{,\theta },
\end{equation}
where $\zeta $ is the ratio between the toroidal field component and the poloidal component.
So the four-vector potential and the magnetic field can be written as:
\begin{equation}
 A(r,\theta )_{\mu }=(0,\zeta e^{\frac{\lambda-\nu  }{2}}a , 0, \sin \theta a_{,\theta } ),
 \end{equation}
\begin{equation}
  B_{\mu }=\frac{e^{\frac{-\lambda }{2}}}{\sin\theta } \bigl(0,\frac{e^{\lambda }(\sin\theta a_{,\theta })_{,\theta }}{r^{2}},-(\sin\theta a_{,\theta })_{,r},-\zeta e^{\frac{\lambda -\nu }{2}}\sin^{2}\theta a_{,\theta } \bigl).              
 \end{equation}.                 

We expand the function $a(r,\theta )$ in Legendre polynomials\cite{bib4}:
 \begin{equation}
      a(r,\theta )=\sum _{l=1}^\infty a_{l}(r)P_{l}(\theta ),
      \end{equation}  
and in Maxwell equation:
 \begin{equation}
J^{\mu }=\frac{1}{4\pi \sqrt{-g}}(\sqrt{-g}F^{\mu \nu })_{,\nu };  
 \end{equation}
 we use the electromagnetic current given by Ref.4:
 \begin{equation}
 4\pi c_{0}(\rho(r) +p(r))r^{2},  
 \end{equation}
where $\rho(r)$ and $p(r)$ are the total energy density and pressure inside our neutron star. 

Finally, after separating variables, we reach the relativistic Grad-Shafranov equation:
 \begin{eqnarray}
 e^{-\lambda (r)}\frac{d^{2}a_{l}}{dr^{2}}+\frac{e^{-\lambda (r)}}{2}\left(\frac{d\nu(r) }{dr}-\frac{d\lambda(r) }{dr} \right)\frac{da_{l}}{dr}+\left( \zeta ^{2}e^{-\nu (r)}-\frac{l(l+1)}{r^{2}} \right)a_{l}=\\ \nonumber
 4\pi c_{0}(\rho(r) +p(r))r^{2}.  
 \end{eqnarray}
 
\section{Solving the equation}
The relativistic Grad-Shafranov equation $(12)$ does not have an analytic solution inside the star and must be solved numerically for each \textit{l} value ($\textit{l}=1$ is the dipole term, $\textit{l}=2$ is the quadrupole term and so on). In our study we focused on the dipole term and varied the ratio $\zeta $, beginning from $\zeta=0$, a pure poloidal field, until $\zeta=0,35$. $\zeta$ can't assume any value, only the fields with continuum domains are allowed in our study, disjoint field lines mean that the $B_{r}$ component changes sign inside the star, but there is no special reason for this change\cite{bib4}.

 Our source term choice implies a discontinuity in the $B_{\phi }$ term in the star's surface where the source term goes to zero. One way to avoid this discontinuity is the introduction of a magnetosphere outside the star, but this is beyond the scope of our work\cite{bib4}.   
 
 In the limit $r\rightarrow  0$ the equation $(13)$ needs to be expanded in powers of $r$:
\begin{eqnarray}
 a_{1}(r\sim 0)=\alpha r^{2}+ \frac{r^{4}}{10}\Big(-\alpha e^{-\nu _{(r=0)}}\zeta ^{2}-\frac{8\pi \alpha }{3}(3p_{(r=0)}-\rho _{(r=0)}) \\ \nonumber
  +4\pi c_{0}(\rho_{(r=0)} +p_{(r=0)})\quad \Big). 
 \end{eqnarray} 
Outside the star, equation $(13)$ has an analytic solution:
\begin{equation}
a_{1}(r)=\frac{-3\mu r^{2}}{8M^{3}}\Bigg[ \ln \Big(1-\frac{2M}{r}\Big)+\frac{2M}{r}+\frac{2M^{2}}{r^{2}} \Bigg],
\end{equation}
and in the limit $r\rightarrow  \infty $ an expansion in Laurent series shows that it takes the form:
\begin{equation}
  \lim_{r\to \infty }a_{1}(r)=\frac{-3\mu}{8M^{3}}\Bigg[\Bigg(-2Mr-2M^{2}-\frac{8M^{3}}{3r}-\frac{4M^{4}}{r^{2}}\Bigg)+2M^{2}+2Mr \Bigg]\longrightarrow 0,
\end{equation}
where we can see that the potential, and consequently the magnetic field, goes to zero as $r$ increases, which is expected from a poloidal field outside the star.

The constants $\alpha $ and $c_{0}$ are found by imposing the continuity of the magnetic field $B_{r}$ and $B_{\theta }$ components at the star surface, $r=R$, and setting the field strength at the star pole, respectively.

The final form of the magnetic field inside the star is:
\begin{equation}
  B_{\mu }= \bigg( 0, \quad \frac{-2e^{\frac{\lambda (r)}{2}}}{r^{2}}a_{1}(r)cos\theta, \quad e^{\frac{-\lambda (r)}{2}}\frac{da_{1}(r)}{dr}sen\theta , \quad \zeta e^{\frac{-\nu (r)}{2}}a_{1}(r)sen^{2}\theta   \bigg).
  \end{equation}
  
\section{Numerical Results} 
 
We solved the equation for the interval $0\leq\zeta  \leq0.35 $, ranging from a pure dipolar field to the first value of $\zeta $ where the first disjointed region occurs. We used a neutron star model with an APR\cite{bib5} equation of state with $1.4M_{\odot}$ and radius of $11.34$km.   

\begin{figure}[t]
\centerline{\includegraphics[width=15cm]{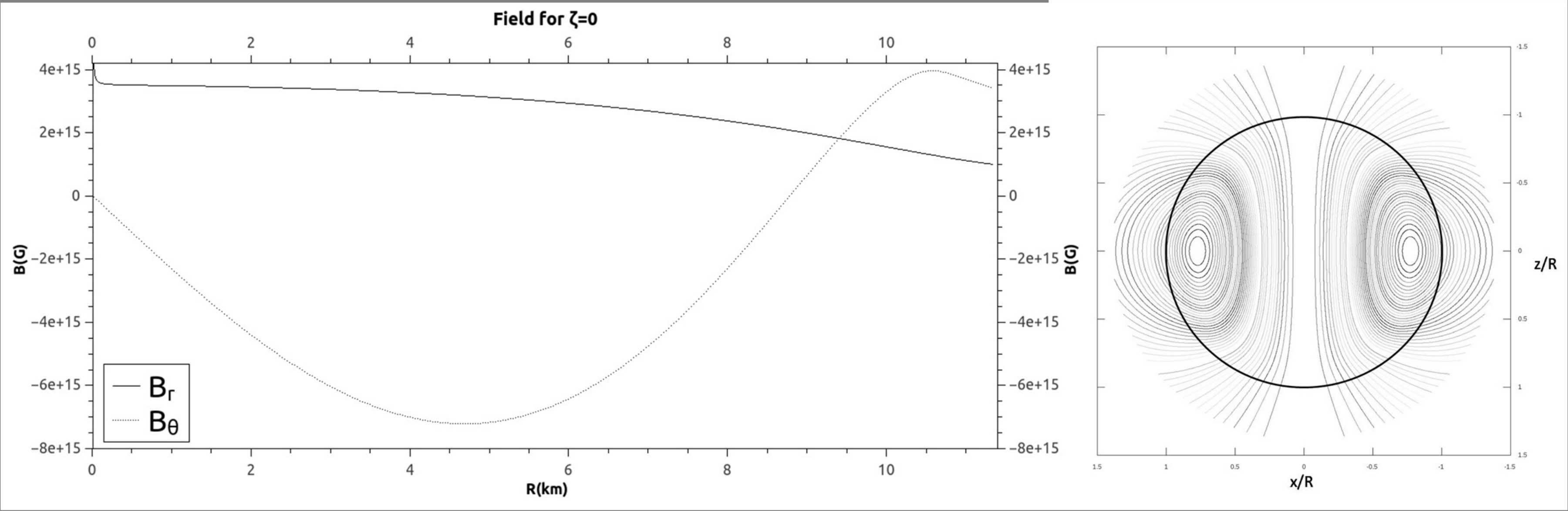}}
\vspace*{8pt}
\caption{The magnetic field for $\zeta =0$, a pure poloidal field. The left side figure shows the field components strength inside the star and the right side figure shows the field line configuration in the star. \label{f1}}
\end{figure}

\begin{figure}[t]
\centerline{\includegraphics[width=15cm]{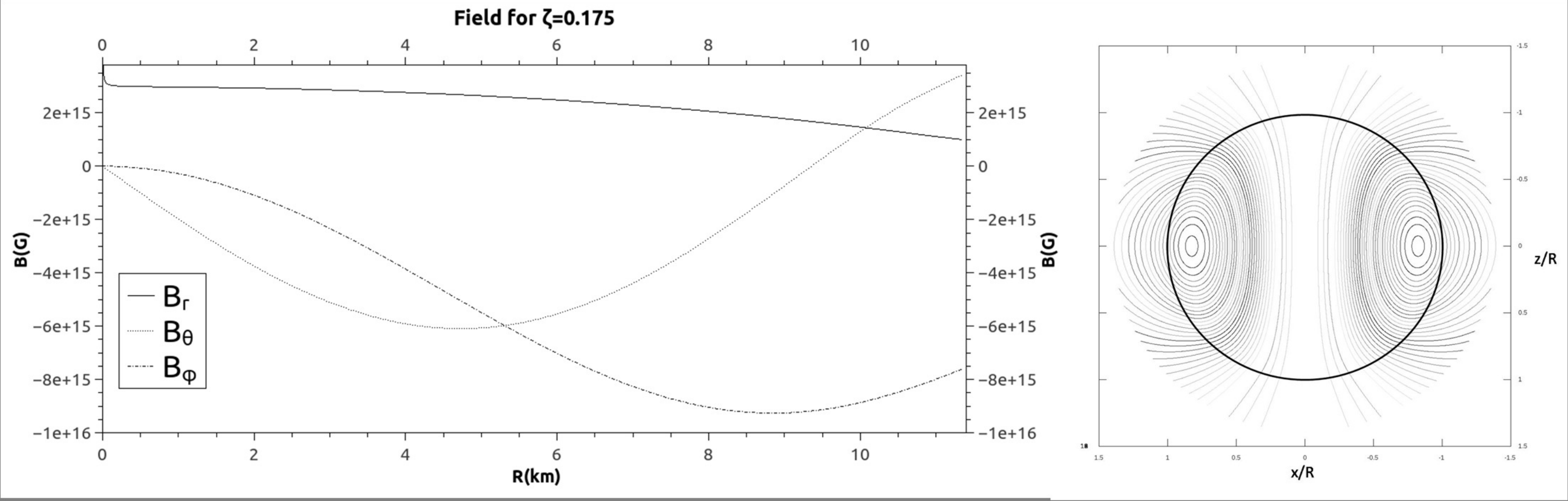}}
\vspace*{8pt}
\caption{The magnetic field for $\zeta =0.175$, an intermediate configuration between the pure poloidal field and the disjoint field configuration. \label{f2}}
\end{figure}

\begin{figure}[t]
\centerline{\includegraphics[width=15cm]{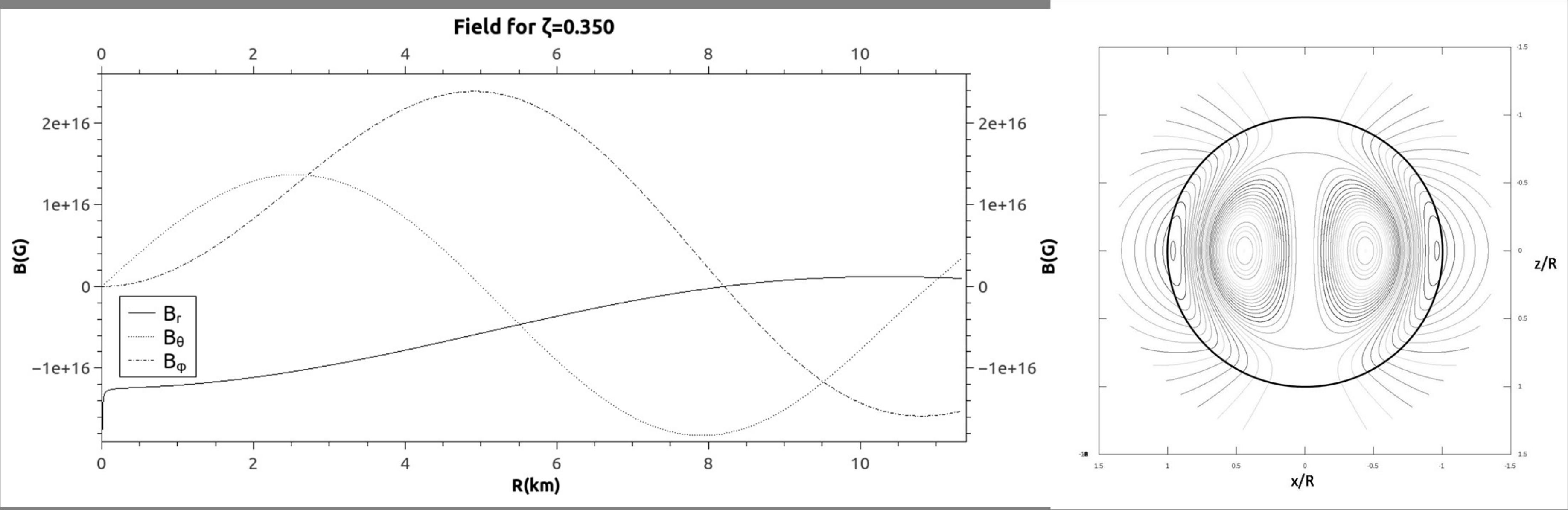}}
\vspace*{8pt}
\caption{The magnetic field for $\zeta =0.350$, the disjoint field configuration. \label{f3}}
\end{figure}

In Fig. 1 we show the components of the magnetic field and its field lines for the case $\zeta =0.00$, the pure dipole case. In all cases we have normalized the field to the value of $10^{15}G$ at the stellar pole. We can see that the $B_{r}$ component does not change sign inside the star and decays softly until it reaches the surface. The $B_{\theta }$ component begins near zero in the origin, increases rapidly in modulus until near 5km and then decreases and reaches zero near 8km, than this component increases again and reaches the surface. The field lines do not have disjointed domains and have a concentration near the surface.

In Fig. 2 we show the components of the magnetic field and its field lines for the case $\zeta =0.175$, an intermediate value between the pure poloidal field and the value of $\zeta =0.350$ . The $B_{r}$ component is similar to the previous case, decaying softly until it reaches the surface. The $B_{\theta }$ does not show a large difference from the previous case. In this case we have the addition of the $B_{\phi }$ component, this component is confined inside the star and we can see that this component's modulus surpasses the other component's modulus. The field lines approach the inner stellar region and tend to form a closed region near the  crust. 

In Fig. 3 we show the components of the magnetic field and its field lines for the case $\zeta =0.350$. The $B_{r}$ component changes sign near 8.5km and this is evident in the field line analysis, where we can see that there are two disjointed region inside the star. The $B_{\theta }$ and $B_{\phi }$ components change the sign twice. According to Ref.4, this happens because the solutions of equation $(13)$ for higher $\zeta $ are linear combinations of the spherical Bessel functions. 

\section{Crustal Oscillations}
As mentioned in section \textbf{1}, \textit{SGR} can be modeled as crust oscillations and three events, named \textit{SGR 1900+14}, \textit{SGR 1802-20} and \textit{SGR 0526-66}, showed a set of distinct periodic oscillation on their data analysis. H. Sotani et al. made a simple model relating the frequencies expected by the model and fitted it to the frequencies measured\cite{bib3}. In their model they restricted the analysis to a dipole magnetic field, without a toroidal component. In our future work we will study the influence of the toroidal component in the crustal oscillations and we will compare the frequencies obtained with those observed.

\section{Conclusions}
Different choices of the $ \zeta  $ parameter allow us to simulate a large range of distinct field configurations. However, a real neutron star with a very strong magnetic field has its shape deformed from the spherical symmetry and this causes tensions on its crust that can literally rip out the crust and free a large amount of energy in the form of gamma rays, this id the proposed mechanism for explaining \textit{SGR} bursts, which we plan to study in our future work.

\section*{Acknowledgments}
We thank \textit{CNPq} (\textit{Conselho Nacional de Desenvolvimento} $Cient\acute{\imath }fico$ \textit{e} $Tecnol\acute{o}gico$) for the financial support and the \textit{IWARA} 2016 organizing committee for the opportunity to attend this event.

\end{document}